\documentclass[english,twoside,a4paper,10pt]{article}

\usepackage[latin1]{inputenc}
\usepackage[T1]{fontenc}
\usepackage{amsmath}
\usepackage{amsfonts}
\usepackage{graphicx}
\usepackage{subfigure}
\usepackage{a4wide}
\usepackage{amssymb}
\usepackage{fancyhdr}
\usepackage{mathrsfs}
\usepackage[toc,page]{appendix}

\begin{document}

\title{Static Spherically Symmetric solutions in a subclass of Horndeski theories of gravity}

\author{ 
Lorenzo Sebastiani\footnote{E-mail address: lorenzo.sebastiani@unitn.it
}\\
\\
\begin{small}
Dipartimento di Fisica, Universit\`a di Trento,Via Sommarive 14, 38123 Povo (TN), Italy
\end{small}\\
\begin{small}
TIFPA - INFN,  Via Sommarive 14, 38123 Povo (TN), Italy
\end{small}
}

\maketitle


\begin{abstract}
In this paper, we will consider a subclass of models of Horndeski theories of gravity and we will check for several Static Spherically Symmetric solutions.
We will find a model which admits an exact black hole solution and we will study its thermodynamics by using the Euclidean Action.
We will see that, in analogy with the case of General Relativity, the integration constant of the solution can be identified with the mass of the black hole itself. 
Other solutions will be discussed,
by posing a special attention on the possibility of reproducing the observed profiles of the rotation curves of galaxies\footnote{Solution (\ref{Bex1})--(\ref{exphi1}) here derived and investigated has been previously discovered by Babichev, Charmousis and Leh\'ebel in Ref.~\cite{Bab}.}.
\end{abstract}



\section{Introduction}

The dark energy issue and the evidences that our Universe underwent a period of strong accelerated expansion after the Big Bang may suggest that a modified theory of gravity different 
to General Relativity (GR)
lies behind our Universe.
The theory of Einstein can be modified in several ways, i.e., by considering a number of dimensions different to four, by introducing additional scalar or vector fields apart the metric
tensor, or
by working with tensor theories which generalize GR
through
the introduction 
of some combinations of the curvature invariants inside the gravitational action (see for instance 
Refs.~\cite{Od, Capo} for the case of $F(R)$-gravity and Ref.~\cite{extgr}). 

If on the one side the introduction of new degrees of freedom in the theory allows to reproduce a huge variety of cosmological scenarios, 
on the other side the field equations risk to become quite involved and 
can be higher derivative. This fact brings to several consequences.  
For example, the theory must be protected against Ostrogradski's instabilities or negative energy states~\cite{Ostro, Ostro2}.
Moreover, despite to the fact that the motivations for alternative theories are mainly cosmological, one cannot ignore the effects of a modification of GR 
at a local scale. 
In the specific, the issue of finding spherical solutions different to the Schwarzshild metric is
a formidable task, since also for a simplified model of modified gravity the equations of motion are much more complicated than those of GR~\cite{SSSsolutions}.
These solutions may exhibit new features not present in GR
(e.g. traversable vacuum wormholes~\cite{WH}, regular black holes, the dark matter
phenomenology) and the formalism of GR may require a new formulation (see for exemple Refs.~\cite{actionFR, FRterm} for the action growth and the 
black hole thermodynamic in tensor theories of modified gravity).

On the other hand, the modified theories of gravity where the field equations are at the second order like in GR deserve a great interest. 
For example, the most general models constructed with the metric tensor and yielding second order differential equations in an arbitrary number of space-time 
dimensions  are known as Lavelock theories~\cite{Lovelock}.

In 1971,  
Horndenski 
found the most general class of scalar-tensor gravitational models where the equations of motion are at the second 
order~\cite{Horn}. The Horndeski Lagrangian is quite involved and includes the Galileian models with Galilean symmetry in flat space-time~\cite{Gal}. 
The Horndenski theories of gravity have been deeply analyzed in literature. They can support the early-time inflation and the theory of perturbations has been fully 
investigated
(for recent works see Refs.~\cite{Def,DeFelice, Kob, Kob2, DeTsu, Qiu, EugeniaH, add0, mioH, add} and references therein).

Local and stationary solutions in Horndeski gravity have been investigated in Ref.~\cite{Rinaldi}, where an exact black hole (BH) solution in the presence of a non minimal coupling
between a scalar field and the Einstein's tensor has been found (see also Refs.~\cite{gen1, gen2} for the generalizations to the case with cosmological constant and 
Refs.~\cite{add1, add2, add3} for the applications of the solution to the neutron stars), and in 
Ref.~\cite{Kob3}, where the authors studied BH solutions in an other subclass of Horndeski theories. In Ref.~\cite{Maselli} Maselli \emph{et al.} furnish the 
equations of motion for spherically symmetric metric in general models of Horndeski theories.

The aim of this paper is to find some new exact spherically symmetric solutions in a subclass of Horndeski gravity and analyze some possible applications 
(black holes, the profiles of the rotation curves of galaxies...).
Moreover, given a black hole solution which is equivalent to the Reissner-Nordstrom solution obtained in the absence of an electric field, 
we investigate its thermodynamics. In GR several thermodynamical definitions are associated to the horizon of a black hole, but in a modified theory 
the metric of a black hole is not expected to
share the same proprieties of its Einsteinian counterparts. When the theory admits higher derivative field equations, several integration constants may emerge from the solution,
and their physical meaning is unclear. In Horndenski gravity the field equations lead to an integration constant in the solutions, and in the case of the BH metric 
such a constant may be interpreted as the energy of the black hole itself.
The energy of a black hole can be defined by making use of the First law of thermodynamics, 
where the entropy may be derived via Wald method~\cite{Wald} or with the euclidea action. How it has been observed in Refs.~\cite{eH1, eH2}, the application of the Wald
formula in the framework of Horndeski gravity could be problematic, since one has to take into account some additional contributions from the scalar field. In our work,
we will work with the Euclidean action with suitable boundary terms~\cite{BTHorn}. We show that the energy of our BH solution is proportional to the integration constant of 
the metric, giving to it a physical meaning. Thus, our theory would be a test bed to reveal some general features of the BH thermodynamics in Horndeski
models of gravity.

The paper is organized in the following way. In Section {\bf 2} the model and the equations of motion for SSS space-time is presented. In Section {\bf 3} we show that the model admits
the Reissner-Nordstrom solution. Since such a solution may describe a black hole, we study its thermodynamics by using the Euclidean action in order to find energy and entropy.
Section {\bf 4} is devoted to the investigation of other possible SSS solutions. Coclusions and final remarks are given in Section {\bf 5}.

We use units of $k_{\mathrm{B}} = c = \hbar = 1$ and set the Planck Mass as
$8\pi /M_{Pl}^2 =1$.

\section{Model and formalism}

The most general scalar-tensor gravitational models where the equations of motion are at the second order like in GR belong 
to the class of Horndeski theories of gravity, whose action (in vacuum) is given by~\cite{Horn}, 
\begin{equation}
I=\int_\mathcal M dx^4\sqrt{-g}\left[\frac{R}{2}+\mathcal L_H\right]\,,\quad \mathcal{L}_H=\sum_{i=2}^5\mathcal{L}_i\,,\label{action0}
\end{equation}
with
\begin{equation}
\mathcal{L}_2=P(\phi,X)\,,\nonumber
\end{equation}
\begin{equation}
\mathcal{L}_3=-G_3(\phi,X)\Box\phi\,,\nonumber
\end{equation}
\begin{equation}
\mathcal{L}_4=G_4(\phi,X)R+G_{4,X}[(\Box\phi)^2-(\nabla_\mu \nabla_\nu \phi)(\nabla^\mu \nabla^\nu \phi)]\,,\nonumber
\end{equation}
\begin{eqnarray}
\mathcal{L}_5&=&G_5(\phi,X)G_{\mu\nu}(\nabla^\mu \nabla^\nu \phi)-\frac{1}{6}G_{5,X}[(\Box\phi)^3-\nonumber\\&&
3(\Box\phi)(\nabla_\mu \nabla_\nu \phi)(\nabla^\mu \nabla^\nu \phi)+2(\nabla^\mu \nabla_\alpha\phi)(\nabla^\alpha \nabla_\beta \phi)(\nabla^\beta \nabla_\mu \phi)]\,.
\label{cGR}
\end{eqnarray}
In Equation (\ref{action0}), $g$ repersents the determinant of the metric tensor $g_{\mu\nu}\equiv g_{\mu\nu}(x^{\mu})$, $\mathcal M$ 
is the space-time manifold, $R$ is the Ricci scalar and corresponds to the Hilbert-Einstein term, while $\mathcal L_H$ includes the 
higher curvature corrections of GR as in (\ref{cGR}), where the scalar field $\phi$ is coupled with gravity. 
Here,
$P(\phi, X)$ and $G_i(\phi, X)$ with $i=3,4,5$ are functions of the scalar field and its kinetic energy, 
\begin{equation}
X=-\frac{g^{\mu\nu}\partial_\mu\phi\partial_\nu\phi}{2}\,.
\end{equation} 
Finally, $\Box=\nabla_\mu\nabla^\mu$ is the d'Alembertian operator associated with the metric, $\nabla_\mu$ being the covariant derivative, 
and $G_{\mu\nu}=R_{\mu\nu}-R g_{\mu\nu}/2$ is the usual Einstein's tensor, where $R_{\mu\nu}$ is the Ricci tensor. 

We will deal with spherically symmetric static (SSS) exact solutions for a subclass of Horndeski models. 
In Ref.~\cite{Rinaldi}, SSS solutions for the Horndeski model with $G_4(\phi, X)=z (X/2)$, $z$ being a constant, has been considered. In such a case, after an integration
by part, a non-minimal coupling between the scalar field and the Einstein's tensor appeared. The model was quite interesting, since on cosmological
background brought to accelerated expansion
without the introduction any scalar potential~\cite{Amendola}.

In our paper we will investigate the case $G_4(\phi, X)\propto\sqrt{|X|}$, namely
\begin{equation}
I=\int_\mathcal M dx^4\sqrt{-g}\left[\frac{R}{2}+P(\phi, X)+\alpha \sqrt{|X|}R+\frac{\alpha}{2\sqrt{|X|}}\left(\frac{|X|}{X}\right)
[(\Box\phi)^2-(\nabla_\mu \nabla_\nu \phi)(\nabla^\mu \nabla^\nu \phi)]
\right]\,,\label{action}
\end{equation}
which corresponds to
\begin{equation}
G_4(\phi, X)=\alpha \sqrt{|X|}\,,\quad
G_3(\phi, X)=G_5(\phi, X)=0\,,
\end{equation}
where $\alpha$ is a constant. 

The choice of the model is strictly connected with the simplified form of the field equations in a SSS background,
where new exact BH solutions or phenomenological solutions 
for the profile of the rotation curves of galaxies
can be found and investigated. 
However, also at the cosmological level, 
this kind of theory may present some interesting features related to the accelerated expansion. In this respect, we mention 
the results of Ref.~\cite{Hviable}, where the viability of Horndeski theories of gravity for the eraly-time inflation  has been carefully investigated. 
The authors found that the perturbations at the end of inflation can coorectly propagate only when
$G_5(\phi, X)\simeq\text{const}$
(in our case, the constant is zero) and $G_4(\phi, X)\rightarrow 0$. It turns out that for large values of $X$ our model falls in this class of theories. 

A general spherically symmetric static solution is described by the metric
\begin{equation}
ds^2=-A(r)dt^2+\frac{dr^2}{B(r)}+r^2\left(d\theta^2+\sin^2\theta d\phi^2\right)\,,
\label{metric}
\end{equation}
where $A(r)$ and $B(r)$ are functions of the radial coordinate $r$. As a consequence, $\phi\equiv\phi(r)$ and $X\equiv X(r)$ such that
\begin{equation}
X=-B(r)\frac{\phi'^2}{2}\,,
\end{equation}
where the prime index denotes the derivative with respect to $r$ and the field is real, namely $0<\phi'^2$. 

The signature of the metric is preserved for $0<B(r)$, which is our natural auumption in order to find exact SSS solutions. 
In this case we can simplify our expessions by removing the moduli, such that $|-B(r)\phi'^2|=B(r)\phi'^2$. On the other hand, 
when the solution ranges in the region $B(r)<0$ (for example, 
inside a BH horizon), the field generally becomes imaginary such that 
$\phi'^2<0$ and our assumption is still valid.

The Equations of motion
(EOMs)  are derived from the action, whose on-shell form for the metric (\ref{metric}) is given by
\begin{eqnarray}
I&=& \int_{\mathcal M} dx^4\left[
\frac{1}{{4
   A(r)^2}}\sqrt{\frac{A(r)}{B(r)}} \left(r^2 B(r) A'(r)^2 \left(\sqrt{2} \alpha 
   \sqrt{B(r)} \phi '(r)+1\right)\right.\right.\nonumber\\&&\left.\left.
   -r A(r) \left(2 \sqrt{2} \alpha  r \sqrt{B(r)}
   A'(r) B'(r) \phi '(r)+r A'(r) B'(r)
   \right.\right.\right.\nonumber\\&&\left.\left.\left.
   +2 \sqrt{2} \alpha  B(r)^{3/2} \left(r
   A''(r) \phi '(r)+A'(r) \left(r \phi ''(r)+4 \phi '(r)\right)\right)+2 B(r)
   \left(r A''(r)+2 A'(r)\right)\right)
   \right.\right.\nonumber\\&&\left.\left.
   -4 A(r)^2 \left(\sqrt{2} \alpha 
   \sqrt{B(r)} \left(2 r B'(r)-1\right) \phi '(r)+r B'(r)+2 \sqrt{2} \alpha 
   B(r)^{3/2} \left(r \phi ''(r)+\phi '(r)\right)
   \right.\right.\right.\nonumber\\&&\left.\left.\left.
   +B(r)-P(\phi, X) r^2-1\right)\right)
\phantom{\frac{0}{0}}\right]\,.\label{completeaction}
\end{eqnarray}
After integration by parts, we are able to recast the lagrangian in a standard form where only the first derivatives of the metric appear, namely,
\begin{equation}
I=\int_\mathcal M dx^4\sqrt{\frac{A(r)}{B(r)}}\left(1+B(r)+B(r)r\left(\frac{A'(r)}{A(r)}\right)+\sqrt{2}\alpha\sqrt{B(r)\phi'(r)^2}+r^2 P(\phi, X)\right)
+I_B
\,.\label{explaction}
\end{equation}
Here,  the boundary term is given by~\cite{BTHorn},
\begin{eqnarray}
I_B&=&
-\sqrt{\frac{A(r)}{B(r)}}\left(\frac{1}{2}+\frac{\alpha\sqrt{B(r)\phi'(r)^2}}{\sqrt{2}}\right)
\left(
4B(r)r+\frac{A'(r)B(r)r^2}{A(r)}
\right)\Big\vert_{\partial_M}\,,\nonumber\\\label{IB}
\end{eqnarray}
where $\partial\mathcal M$ denotes the surface of the manifold $\mathcal M$.
In the case of GR with $\alpha=0$ one recovers the Gibbons-York-Hawking boundary term~\cite{Gibbons}. The boundary term does not contribute to the field 
equations of the theory and one can work with the bulk action in (\ref{explaction}) only.
Thus,
the variations with respect to the metric functions $A(r), B(r)$ lead to
(see also Ref.~\cite{Maselli}),
\begin{equation}
1-B(r)-r B'(r)+\sqrt{2}\alpha\sqrt{B(r)\phi'(r)^2}=-r^2 P(\phi, X)\,,\label{9}
\end{equation}
\begin{equation}
-1+B(r)\left(1+\frac{rA'(r)}{A(r)}\right)=r^2 P(\phi, X)+r^2 B(r)\phi'^2 P_X(\phi, X)\,.
\label{10}
\end{equation}
Moreover, the variation with respect to the field reads
\begin{eqnarray}
&& P_\phi(\phi,X)+\frac{2B(r)}{r}\phi' P_X(\phi,X)+\frac{B'(r)\phi'}{2}P_X(\phi,X)+B\phi'' P_X(\phi,X)
\nonumber\\&&
+\frac{A'}{2A\phi'}\left[
-\frac{\alpha\sqrt{2B(r)\phi'^2}}{r^2}+B\phi'^2 P_X(\phi,X)
\right]+B\phi'^2 P_{X\phi}(\phi, X)-\frac{B B'\phi'^3P_{XX}(\phi, X)}{2}\nonumber\\&&
-B^2\phi'^2\phi'' P_{XX}(\phi, X)=0\,.
\label{11}
\end{eqnarray}
In the next sections, we will check for some explicit, exact solutions of the model.

\section{An exact black hole solution and its thermodynamics}

We will consider the case of a canonical scalar field which is not subject to any potential, namely
\begin{equation}
P(\phi, X)=X\,.\label{exone}
\end{equation}
From Eq.~(\ref{10}) we have
\begin{equation}
\phi'(r)=\pm\frac{\sqrt{-2 A(r)+2A(r)B(r)+2B(r)A'(r)r}}{\sqrt{A(r)B(r)}r}\,.
\end{equation}
Thus, Equation (\ref{9}) reads
\begin{equation}
B(r) \frac{A'(r)}{A(r)}+\frac{1}{r}\left(
-2+2B(r)-\frac{2\alpha}{r}\sqrt{\frac{A(r)(B(r)-1)+r B(r)A'(r)}{A(r)}}
+r B'(r)
\right)=0\,,
\end{equation}
and the implicit form of $A(r)$ is derived as
\begin{equation}
A(r)=A_0 \text{Exp}\left[
\int dr
\frac{-2B(r) r^2-r^3 B'(r)+2\left(
\alpha^2+r^2\pm\alpha\sqrt{\alpha^2+r^2-B(r)r^2-r^3 B'(r)}
\right)}
{B(r)r^3}
\right]\,.\label{A1}
\end{equation}
In what follows, we will pose $A_0=1$. The discriminant in (\ref{A1})
can be eliminated by 
choosing
\begin{equation}
B(r)=1-\frac{M}{r}-\frac{\alpha^2}{r^2}\,,\label{Bex1}
\end{equation}
where $M$ is a mass constant. It follows
\begin{equation}
A(r)=B(r)\,,\label{Aex2}
\end{equation}
and also Equation (\ref{11}) is satisfied. 
What we have obtained is a Reissner-Nordstrom solution, where the role of the electric charge is played by the coupling 
constant $\alpha$ between the field and gravity. In the limit $\alpha=0$ we correctly recover the Schwarzshild solution of GR. 
The explicit form of the field is given by
\begin{eqnarray}
\phi(r)&=&\phi_0\pm\sqrt{2}\arctan\left[
\frac{2\alpha^2+M r}{2\alpha\sqrt{-M r+r^2-\alpha^2}}
\right]\,,\quad \alpha\neq 0\,,\nonumber\\
\phi(r)&=&\phi_0\,,\quad \alpha=0\,,\label{exphi1}
\end{eqnarray}
such that
\begin{eqnarray}
\phi'(r)&=&\pm
\frac{\sqrt{2}\alpha}{r\sqrt{r^2-M r-\alpha^2}}\,,
\quad \alpha\neq 0\,,\nonumber\\
\phi'(r)&=&0\,,\quad \alpha=0\,,\label{exphi1bis}
\end{eqnarray}
with $\phi_0$ a generic integration constant.
We should note that when we take $\sqrt{\phi'(r)^2}=|\phi'|$ in (\ref{9}) and (\ref{11}), such equations are consistent with the given solution only for $0<\alpha$.

The field is real when the metric signature is preserved, namely (for $r$ real),
\begin{equation}
\frac{M+\sqrt{M^2+4\alpha^2}}{2}<r\,.
\end{equation}
The only positive root of $B(r)$ is located at $r=r_H$ such that
\begin{equation}
r_H=\frac{M+\sqrt{M^2+4\alpha^2}}{2}\,,
\end{equation}
and, since $0<B'(r_H)$, we are in the presence of the event horizon of a black hole.\\
\\
To the event horizon of a static black hole is possible to associate a Killing surface gravity $\kappa_K$ and therefore a Killing temperature $T_K$, 
and for the metric (\ref{metric}) we have~\cite{HT},
\begin{equation}
T_K:=\frac{\kappa_K}{2\pi}=\sqrt{\frac{B(r_H)}{A(r_H)}}\frac{A'(r_H)}{4\pi}\,.\label{HawkingTemperature}
\end{equation}
Here, a remark is in order. The generalization of the formalism to the dynamical case requires a covariant formulation where
the Kodama vector field~\cite{Kodama} replaces the time-like Killing vector field. When we come back to the stationary case,
due to the different normalization of the two vectors, we have two definitions for the surface gravity and for the BH temperature.
For the metric (\ref{metric}) the Kodama/Hayward temperature $T_H$ reads~\cite{Hay},
\begin{equation}
T_H:=\frac{\kappa_H}{2\pi}=\frac{1}{4\pi}\left[
\frac{B'(r)}{2}+\frac{A'(r) B(r)}{2A(r)}
\right]_H\,,
\end{equation}
where $\kappa_H$ is the Hayward surface gravity. For the Reissner-Nordstrom solution (\ref{Bex1})--(\ref{Aex2}), due to the fact that $A(r)=B(r)$, no ambiguities exist and the 
BH temperature is well defined by (\ref{HawkingTemperature}),
\begin{equation}
T_K=\frac{r_H^2+\alpha^2}{r_H^3}\,.\label{TRN}
\end{equation}
such that $T_K=T_H$.

To evaluate the entropy of a black hole, one way is to pass to the Euclidean action $I_E$ by redefining the 
time as $t\rightarrow i\tau$ inside the classical action. For an useful and complete discussion about the entropy issue 
in Horndeski gravity and the corrections to the Wald formalism with different approaches, see Refs.~\cite{eH1,eH2}.

Given the ($\tau, r$)-components of the euclidean metric near to the horizon,
\begin{equation}
ds^{2(2)}_E=\left(\frac{B(r_H)}{A(r_H)}\frac{A'(r_H)^2}{4}\right)\sigma^2 d\tau^2+d\sigma^2\,,\quad\sigma=2\sqrt{\frac{(r-r_H)}{B'(r_H)}}\,,
\end{equation}
the conical singularity is removed if $\tau\in [0, \beta]$, where the period $\beta$ correponds to
\begin{equation}
\beta=\frac{2\pi}{\kappa_K}\equiv\frac{1}{T_K}\,,\label{beta}
\end{equation}
and it is related to the Killing temperature of the black hole itself. Thus, the Euclidean action reads
\begin{equation}
I_E=-4\pi\beta\left[\int_{r_H}^\infty dr\mathcal L-I_B |_{r\rightarrow\infty} \right]\,,\label{IE}
\end{equation}
where $\mathcal L$ is the standard Lagrangian in (\ref{completeaction}) and $I_B$ is the boundary term in (\ref{IB}) and must be subtracted in order to have a well posed variational principle
working with the bulk action in (\ref{explaction}) only. When one passes to the Euclidean time, the surface of the manifold gives a contribution only for $r\rightarrow\infty$.
Moreover, some additional suitable counterterms may be necessary to regularize the action (see Refs.~\cite{revBH1, revBH2} for some reviews).

Once the Euclidean action is given, one can derive the energy $E$ and the entropy $S$ of the black hole by making use of the following relations,
\begin{equation}
M=\frac{\partial I_E}{\partial\beta}\,,\quad S=\beta\frac{\partial I_E}{\partial\beta}-I_E\,,
\end{equation}
such that the First law of thermodynamics holds true as,
\begin{equation}
T_K dS=dE\,.\label{FL}
\end{equation} 
Let us return to our model in (\ref{action}) with (\ref{exone}). Given the BH solution (\ref{Bex1})--(\ref{Aex2}),
the contribution from the boundary term in (\ref{IE}) reads,
\begin{equation}
4\pi\beta I_B |_{r\rightarrow\infty}=
4\pi\beta\left(
-2r-\frac{3\alpha^2}{r}+\frac{3M\alpha^2}{r^2}+\frac{2\alpha^4}{r^3}+\frac{3M}{2}
\right)|_{r\rightarrow\infty}=4\pi\beta\left(
-2r+\frac{3M}{2}
\right)|_{r\rightarrow\infty}\,.\label{Bound0}
\end{equation}
The divergence can be eluiminated by subracting the boundary term with integration constant $M=0$~\cite{Page} and a period $\beta_2$ identified with the inverse
temperature of the $M=0$ background. We require that, for large values of $r$, the metric components $g_{\tau\tau}$ of the euclidean metrics with $M\neq 0$ and $M=0$ are equal, namely  
\begin{equation}
\beta_2=\beta \sqrt{\frac{B(M\neq 0, r\rightarrow \infty)}{B(M=0, r\rightarrow\infty)}}\simeq\beta\left(1-\frac{M}{2r}\right)|_{r\rightarrow\infty}\,. 
\end{equation}
Thus, the contribution of the boundary in (\ref{Bound0}) is regularized as
\begin{equation}
I_B=4\pi\beta\left(
-2r+\frac{3M}{2}+2r\left(1-\frac{M}{2r}\right)
\right)|_{r\rightarrow\infty}=4\pi\beta\frac{M}{2}\,.\label{Bound}
\end{equation}
We finally obtain for the Euclidean action (\ref{IE}),
\begin{equation}
I_E= 4\pi\beta\alpha^2\left[
\frac{3M}{r_H^2}-\frac{5}{r_H}+\frac{2\alpha^2}{r_H^3}+\frac{M}{2\alpha^2}
\right]\,,
\end{equation}
and by taking into account (\ref{TRN}, \ref{beta}) togheter with the horizon condition $M=r_H-\alpha^2/r_H$, we get
\begin{equation}
E=4\pi M\,,\quad S=(4\pi)^2\left(\alpha^2+\frac{r_H^2}{2}\right)\,.
\end{equation}
We remember that the results are expressed in units of $8\pi/M_{Pl}^2$. In our model, the energy of a Reissner-Nordstrom black hole is the same of the one in the background of GR, where 
the First law in (\ref{FL}) aquires an addition contribution from the pressure of the electromagnetic field. On the other hand, the BH entropy 
is larger than the
Area law formula predicrted by the Hawking radiation in the framework of GR.

\section{Other SSS solutions}

In this section, we will derive other exact SSS solutions of our Lagrangian in (\ref{action})
for different choices of $P(\phi, X)$. Let us start by taking the non-canonical form 
\begin{equation}
P(\phi, X)=\lambda\sqrt{|X|}\,,
\end{equation}
where $\lambda$ is a generic positive dimensional constant. 
Equations (\ref{9}, \ref{10}) yield
\begin{equation}
2-2B(r)-2r B'(r)+2\alpha\sqrt{2B(r)\phi'(r)^2}+r^2\lambda\sqrt{2B(r)\phi'(r)^2}=0\,,\label{9caso2}
\end{equation}
\begin{equation}
1-B(r)-rB(r)\frac{A'(r)}{A(r)}=0\,,
\end{equation}
with the implicit solutions
\begin{equation}
\phi'(r)=\pm\frac{\sqrt{2}(-1+B(r)+r B'(r))}{\sqrt{B(r)(4\alpha^2+4r^2\alpha\lambda+r^4\lambda^2)}}\,,
\end{equation}
\begin{equation}
A(r)=A_0\text{Exp}
\left[\int
\frac{(1-B(r))}{B(r)r}dr
\right]\,,\quad \alpha\,,\lambda\neq0\,.
\end{equation}
Here, $A_0$ must be a positive constant and will be fixed as $A_0=1$. 
If we plug the expressions above in (\ref{11}), we obtain
\begin{equation}
\left(-2\alpha-r^2\lambda+(2\alpha-3r^2\lambda)B(r)
\right)=0\,,
\end{equation}
which leads to
\begin{equation}
B(r)=\frac{2\alpha+r^2\lambda}{2\alpha-3r^2\lambda}\,,
\end{equation}
such that,
\begin{equation}
A(r)=\frac{1}{(2\alpha+r^2\lambda)^2}\,,\quad \phi(r)=\pm\int
\frac{12\sqrt{2}r^2\lambda(2\alpha-\lambda r^2)}{(2\alpha-3r^2\lambda)^{3/2}}\frac{\sqrt{2}}{\sqrt{(2\alpha+\lambda r^2)}
|2\alpha+\lambda r^2|}dr\,,\quad \lambda\neq 0\,.
\end{equation}
In this solution any integration constant emerges and the parameter $\lambda$ is in fact a cosmological constant which causes the appearence 
of a cosmological horizon. In the limit $\lambda\rightarrow 0$ (eventually, with $\alpha=1/2$) we get the Minkowski space-time. If 
$0<\alpha\,,\lambda$, the metric
signature is preserved as long as $-2\alpha<\lambda r^2<2\alpha/3$. On the other hand, if $\alpha\,,\lambda<0$,
the metric signature is preserved when $2\alpha/3<\lambda r^2<-2\alpha$.\\
\\
Let us introduce now a potential for the field in the following way,
\begin{equation}
P(X,\phi)=-V(\phi)\,.\label{-V}
\end{equation}
Thus, on shell, the potential can be treated as a function of $r$ such that
\begin{equation}
V(\phi)=V(r)\,,\quad V_\phi(\phi)=\frac{V'(r)}{\phi'}\,,
\end{equation}
and the Equations (\ref{9},\ref{10},\ref{11}) read,
\begin{equation}
-1+B(r)+r^2 V(r)+r B'(r)-\alpha\sqrt{2B(r)\phi'^2}=0\,,\label{unobis}
\end{equation}
\begin{equation}
A(r)\left(-1+B(r)+r^2 V(r)\right)+r B(r) A'(r)=0\,,\label{duebis}
\end{equation}
\begin{equation}
2V'(r)+\frac{\alpha A'(r)\sqrt{2 B(r)\phi'^2}}{r^2A(r)}=0\,.\label{trebis}
\end{equation}
From Eq.~(\ref{duebis}) we directly obtain
\begin{equation}
V(r)=\frac{A(r)-A(r)B(r)-r B(r)A'(r)}{r^2 A(r)}\,,
\end{equation}
while Eq.~(\ref{unobis}) leads to
\begin{equation}
\phi'(r)=\pm\frac{B(r)r A'(r)-A(r) r B'(r)}{\alpha A(r)\sqrt{2B(r)}}\,.
\end{equation}
Now, the implicit solution of $B(r)$ is derived from Eq.~(\ref{trebis}).  If $B(r)\neq A(r)$ one has,
\begin{eqnarray}
B(r)&=&c_0 \exp \left(\int^r \frac{-2 \left(r'\right)^2 A\left(r'\right) A''\left(r'\right)+2 r'
   A'\left(r'\right) A\left(r'\right)+\left(r'\right)^2 A'\left(r'\right)^2+4 A\left(r'\right)^2}{r'
   A\left(r'\right) \left(r' A'\left(r'\right)+2 A\left(r'\right)\right)} \, dr'\right)+\nonumber\\
&&\exp
   \left(\int^r \frac{-2 \left(r'\right)^2 A\left(r'\right) A''\left(r'\right)+2 r'
   A'\left(r'\right) A\left(r'\right)+\left(r'\right)^2 A'\left(r'\right)^2+4 A\left(r'\right)^2}{r'
   A\left(r'\right) \left(r' A'\left(r'\right)+2 A\left(r'\right)\right)} \, dr'\right)\times\nonumber\\
&&\hspace{-1cm}
 \int^r
   -\frac{4 A\left(r''\right) 
\exp \left(-\int^{r''} \frac{-2 \left(r'\right)^2 A\left(r'\right)
   A''\left(r'\right)+2 r' A'\left(r'\right) A\left(r'\right)+\left(r'\right)^2 A'\left(r'\right)^2+4
   A\left(r'\right)^2}{r' A\left(r'\right) \left(r' A'\left(r'\right)+2 A\left(r'\right)\right)} \,
   dr'\right)}{r'' \left(r'' A'\left(r''\right)+2 A\left(r''\right)\right)} \, dr''\,,
\end{eqnarray}
where $c_0$ is an integration constant. When $B(r)=A(r)$ the field results 
to be a constant and $V(\phi)=0$, such that one recovers the Schwarzshild solution of General Reltivity.

We will look for some exact solutions. If we take the Newtonian form for $A(r)$,
\begin{equation}
A(r)=A_0\left(1-\frac{M}{r}\right)\,,
\end{equation}
where $M$ is a mass constant and, as usually, we put $A_0=1$, 
we obtain
\begin{equation}
B(r)=\left(1-\frac{M}{r}\right)\left(1+c_0(M-2r)^2\right)\,.
\end{equation}
Here, $c_0$ is a fixed parameter of the model which appears in the potential. One easily reconstructs,
\begin{equation}
V(r)=-\frac{c_0(M-2r)^2}{r^2}\,,\quad\phi'(r)=\pm c_0\frac{2\sqrt{2}(M-2r)(M-r)}{\alpha\sqrt{B(r)}}\,.
\end{equation}
Thus, as long as
\begin{equation}
M<r\,,
\end{equation}
the metric signature is preserved and the field is real.

The explicit form of the potential can be derived only in the limiting cases $c_0=0$ or $M=0$.
For $c_0=0$ one recovers the Schwarzshild solution. For $M=0$ one has,
\begin{equation}
\phi(r)=\phi_0\pm\frac{\text{arcsinh}[2r\sqrt{c_0}]-2r\sqrt{c_0}\sqrt{1+4r^2 c_0}}{2\alpha\sqrt{2c_0}}\,,\quad 0<c_0\,,
\end{equation}
\begin{equation}
\phi(r)=\phi_0\pm\frac{\text{arcsin}[i 2r\sqrt{-c_0}]-2r\sqrt{-c_0}\sqrt{1+4r^2 c_0}}{2\alpha\sqrt{-2c_0}}\,,\quad c_0<0\,,
\end{equation}
$\phi_0$ being a constant, and
\begin{equation}
V(r)=-4c_0\,.
\end{equation}
In this case, according with (\ref{-V}), the potential plays the role of a negative/positive cosmological constant, and the $g_{rr}$ component 
of the metric takes the Anti-de Sitter/de Sitter (AdS/dS) form, while $g_{tt}$ is equal to 
$-1$ like in the flat space-time\footnote{The model exhibts also an implicit solution where $B(r)=1$. In such a case one obtains $A(r)=(1+\gamma r)^2$, $\gamma$ being a constant.}.
We mention that, given $\sqrt{\phi'^2}=|\phi'|$, the equations (\ref{unobis}), (\ref{trebis}) are consistent only when $0<\alpha$.

Other interesting solutions can be found for the following class of Lifschitz-like solutions,
\begin{equation}
A(r)=\left(\frac{r}{r_0}\right)^z B(r)\,,\label{AA}
\end{equation}
where $r_0$ is a lenght scale and $z$ is a number. In this case, the function $B(r)$ assumes the form,
\begin{equation}
B(r)=\frac{4}{4+4z-z^2}+\frac{c_{1,2}}{r^{b_{1,2}}}\,,\quad b_{1,2}=
\frac{1}{4}\left(3z-2\pm\sqrt{\frac{(2 + z) (18 + z) (-4 + ( z-4) z)}{-4 + (z-4) z}}\right)\,,
\end{equation}
where $c_{1,2}$ are two integration constants. We should note that, since the theory admits first order differential equations, 
one integration constant of the solution will be fixed by the Lagrangian of the model.
When $z=0$ such that $b_{1,2}=1,-2$, we recover the Schwarzshild AdS/dS solution. Other BH solutions can be found for positive values of $b_{1,2}$.

In the context of the study of the rotation curves of galaxies, one may consider the following behaviour of $A(r)$,
\begin{equation}
A(r)= \left(1-\frac{M}{r}\right)(1+\gamma r)^2\,,\label{Arot}
\end{equation}
$M\,,\gamma$ being positive constants,
which leads to,
\begin{equation}
B(r)= \frac{4 (M-r) \left(\gamma  M^2 (5 \gamma  r+4)+M \left(5 \gamma ^2
   r^2+2 \gamma  r-2\right)+2 \gamma  r^2\right)}{M r (5 \gamma  M-4)
   (7 \gamma  M-2)}\,.
\end{equation}
From (\ref{Arot}) we can derive the Newtonian potential, 
\begin{equation}
\Phi:=-\frac{(g_{tt}+1)}{2}=\gamma r+\frac{\gamma^2 r^2}{2}-\frac{M}{2r}\left(1+2r\gamma+r^2\gamma^2\right)\,. 
\end{equation}
Thus, we can study the solution in the range $\gamma r/2\ll 1$ and $1\ll \gamma r/2$. If $2/\gamma$ is the typical galactic lenght scale, at the cosmological level $1\ll \gamma r/2$,
when one neglect the contribute of $M$,
a cosmological constant term $\gamma^2 r^2$ emerges in the solution and has an AdS-like form, while, more interesting, at the galactic scale $\gamma r/2\ll 1$ the Newtonian potential 
grows up linearly with the distance as,
\begin{equation}
\Phi\simeq
-\frac{M}{2r}\left(
1-\frac{2\gamma r^2}{M}
\right)
\,. 
\end{equation}
Correspondingly, the rotational velocity profile of the galaxies is derived as,
\begin{equation}
v^2\simeq v_N^2+\gamma c^2 r\,, 
\end{equation}
where we have introduced the light speed $c$ and $v_N$ is the contribution expected from the matter component and decreases as $\sim 1/\sqrt{r}$.
It means that , on sufficiently
large scales, the rotational velocity does not fall-off due to the
Keplerian result $\sim 1/\sqrt{r}$, but increases slightly as $\sim\sqrt{r}$ according with observations~\cite{Riegert, Mannheim, dwarf, capo, salucci}. 
Therefore, Horndeski gravity may be a good framework for the dark matter phenomenology at the galactic scales. 

We should note that he reconstruction of the potential can be done only on shell and its explicit form can be found only in the limiting cases. 
What one will find is that the form of the potential
fixes the scale $\gamma$, while $M$ remains the only free integration constant of the solution.


 
\section{Conclusions}

In this paper, motivated by a general interest in Horndeski theories of gravity, we tried to carried out an analysis on static spherically 
symmetric solutions for a subclass of Horndeski models. We should mention that Horndeski gravity is mainly used to reproduce some cosmological features of our Universe, 
in particular the early-time inflation. The recent paper in Ref.~\cite{Hviable} demonstrated that a linear coupling between the kinetic term 
of the Horndeski scalar field and the Einstein's tensor does not allow for a correct propagation of the primordial cosmological perturbations. Here, we analyzed a model where the Einstein's tensor 
is coupled with the square of the kinetic term of the field which is not exluded \textit{a priori} by observations. 

If one believes that a modified gravity theory governs our Universe, it is important to know how the definitions and the laws of Einstein's 
gravity can be generalized in the modified gravity 
framework. In this respect, the physics of black holes represents an interesting field of research. 
In our class of Horndeski gravitational models we have found a Lagrangian which admits the
Reissner-Nordstrom metric as a vacuum solution. In order to study the thermodynamics of the related black hole, 
we have used the Euclidean action. In this way, by using the First Law of thermodynamics,
we were able to infer the energy and the entropy of our BH solution. Respect to the case of GR, we have found that the entropy of the
black hole is larger than the entropy predicted
by the Area Law. However, the energy turns out to be the same as in Einstein's gravity, namely it 
can be identified with the integration constant of the solution. 

In the last part of the work we have further investigate other possible SSS solutions for our model. We showed that Horndeski gravity may 
produce the correct predictions for the spectra of the rotation curves of galaxies.

\end{document}